\documentclass[%
reprint,superscriptaddress, amsmath,amssymb,aps,prl,linenumbers,
floatfix,
]{revtex4-1}
\usepackage{graphicx}% Include figure files
\usepackage{dcolumn}% Align table columns on decimal point
\usepackage{bm}% bold math
\usepackage{hyperref}% add hypertext capabilities
\hypersetup{colorlinks=true, citecolor=blue, urlcolor=blue, linkcolor=blue}
\usepackage{braket}
\usepackage{accents}
\usepackage{multirow}
\usepackage{dcolumn,booktabs}
\newcolumntype{d}[1]{D{.}{.}{#1}}

\newcommand\mc[1]{\multicolumn{1}{c}{#1}}

\begin{document}

\preprint{APS/123-QED}

\title{Hubbard \textit{\textbf{U}} through  polaronic defect states}

\author{Stefano Falletta}
\email{stefano.falletta@epfl.ch}
\affiliation{Chaire de Simulation \`a l'Echelle Atomique (CSEA), Ecole Polytechnique F\'ed\'erale de Lausanne (EPFL), CH-1015 Lausanne, Switzerland}
\author{Alfredo Pasquarello}
\affiliation{Chaire de Simulation \`a l'Echelle Atomique (CSEA), Ecole Polytechnique F\'ed\'erale de Lausanne (EPFL), CH-1015 Lausanne, Switzerland}

\date{\today}
\nolinenumbers

\begin{abstract}
	\setlength{\leftmargin}{-1mm}
\textbf{Since the preliminary work of Anisimov and co-workers, the Hubbard corrected DFT+\textit{\textbf{U}} functional has been used for predicting properties of correlated materials by applying on-site effective Coulomb interactions to specific orbitals. However, the determination of the Hubbard \textit{\textbf{U}} parameter has remained under intense discussion despite the multitude of approaches proposed. Here, we define a selection criterion based on the use of polaronic defect states for the enforcement of the piecewise linearity of the total energy upon electron occupation. 
A good agreement  with results from piecewise linear hybrid functionals is found for the  electronic and structural properties of polarons, including the formation energies. The values of \textit{\textbf{U}} determined in this way are found to give a robust description of the polaron energetics upon variation of the considered state. In particular, we also address  a polaron hopping pathway, finding that the determined value of \textit{\textbf{U}} leads to accurate energetics without requiring  a configurational-dependent \textit{\textbf{U}}. It is emphasized that the selection of \textit{\textbf{U}} should be based on physical properties directly associated with the orbitals to which \textit{\textbf{U}} is applied, rather than on more global properties such as band gaps and band widths.  For comparison, we also determine \textit{\textbf{U}} through a well-established linear-response scheme finding noticeably different values of \textit{\textbf{U}} and consequently different formation energies. Possible origins of these discrepancies are discussed. As case studies, we consider the self-trapped electron in BiVO$_4$, the self-trapped hole in MgO, the Li-trapped hole in MgO, and the Al-trapped 
hole in $\alpha$-SiO$_2$. }
\end{abstract}

\maketitle

%\linenumbers

\section{Introduction}

Density functional theory (DFT) including a Hubbard $U$ correction has been largely used to overcome limitations of standard DFT for correlated systems \cite{anisimov1991PRB,anisimov1991PRB2,anisimov1993PRB,solovyev1994PRB,czyifmmode1994PRB,liechtenstein1995PRB,anisimov1997JPCM,dudarev1998PRB,petukhov2003PRB,cococcioni2005PRB}. However, the parameter $U$ associated with an effective on-site Coulomb interaction on selected orbitals needs to be selected. In 2005, Cococcioni and de Gironcoli introduced a nonempirical linear-response approach  based on density-functional perturbation theory  \cite{cococcioni2005PRB}, which has largely been applied \cite{zhou2004PRB,tao2014PRB,himmetoglu2014JQC,bjaalie2015PRB,mann2016JCP,ricca2019PRB,timrov2018PRB,floris2020PRB,timrow2021PRB,timrov2022CPC}. In other studies, the parameter $U$ is  chosen to reproduce specific experimental properties, such as band gaps \cite{deskins2007PRB,dudarev2019PRM}, reaction enthalpies \cite{wang2006PRB,garciamota2012JCP,jain2011PRB},  oxidation energies \cite{bajdich2013JACS}, activation energies \cite{deskins2007PRB}, atomic structures \cite{franchini2007PRB}, density of states \cite{aschauer2013PRB}, or magnetic arrangements \cite{hong2012PRB}. 
Alternative strategies consist in fixing $U$ to yield states in the middle of the band gap \cite{deskins2009JPCC}, to comply with criteria based on energy barriers \cite{deskins2009JPCC2}, to have vanishing quasiparticle corrections to the fundamental band gap \cite{patrick2012JPCM}, or to match hybrid-functional results \cite{erhart2014PRB}. The parameter $U$ has also been calculated through  an alternative linear-response method \cite{kulik2006PRL}, through unrestricted Hartree-Fock approach \cite{mosey2007PRB,mosey2008JCP},  through the random-phase approximation  \cite{aryasetiawan2006PRB,miyake2008PRB,sasioglu2011PRB,setvin2014PRL},  through Monte Carlo sampling \cite{tavadze2021NCM}, and through machine-learning techniques based on Bayesian optimization \cite{yu2020NCM}. Clearly, a general consensus on the way $U$ should be determined is still lacking.

In the linear-response approach of Cococcioni and de Gironcoli \cite{cococcioni2005PRB}, $U$ is fixed to comply with the piecewise linearity condition (PWL) of the total energy upon electron occupation, which  is a property of the exact density functional \cite{perdew1982PRL,ruzsinszky2007JCP,zhang1998JPC,yang2000PRL,morisanchez2006JCP}. While being defined for fractional charges,  the PWL allows for an accurate description of ground state and excited state properties of systems with integer number of electrons \cite{kronik2020PCCP}. 	Most density functionals do not comply with the PWL. For instance, the total energy obtained with the Perdew-Burke-Ernzerhof (PBE) \cite{perdew1996PRL} semilocal functional is convex with the number of electrons. Similarly, the total energy obtained with the Hartree-Fock functional is concave. However, the PWL can be retrieved through suitably tuned functionals.   For instance, for hybrid functionals \cite{perdew1996JCP}, there exists a fraction $\alpha=\alpha_\text{k}$ of Fock exchange for which the total energy is linear upon electron occupation. Through Janak's theorem \cite{janak1978PRB}, this results in a generalized Kohn-Sham level that is constant upon electron occupation. Additionally, under this condition, band gaps and formation energies of localized states are accurately reproduced \cite{miceli2018PRB,deak2017PRB,kronik2020PCCP,sadigh2015PRB,sai2011PRL,refaely2013PRB,bischoff2019PRM,bischoff2019PRB,bischoff2021PRR,yang2022JPCL,falletta2022PRL,falletta2022PRB}.

Localized states represent a prototypical case for enforcing the PWL. For instance, for hybrid functionals, this can be achieved by using either electron probes \cite{bischoff2019PRM,bischoff2019PRB,bischoff2021PRR,yang2022JPCL}, defect states \cite{miceli2018PRB,peng2017PNAS,sadigh2015PRB,kokott2018NJP,ambrosio2018EES,elmaslmane2018JCTC,carey2019JPCC,carey2021JPCC,falletta2020PRB,osterbacka2020CM,quirk2020JPCC,falletta2022PRL,falletta2022PRB}, or Wannier functions \cite{wing2021PNAS}. In the context of polarons, the PWL has been used to regulate the strength of potentials added to the semilocal Hamiltonian to favor charge localization, as in the schemes of Lany and Zunger \cite{lany2009PRB} and of Falletta and Pasquarello \cite{falletta2022PRL,falletta2022PRB}. Moreover, the properties of polaronic defects are found to be robust for semilocal or hybrid functionals complying with the PWL  \cite{falletta2022PRL,falletta2022PRB}.  
Hence, it is of interest to investigate whether such robustness can be used to validate the determination of $U$ in  DFT+$U$ functionals.

In this work, we determine the Hubbard parameter $U$ by using polaronic defect states to explicitly enforce the piecewise linearity of the total energy upon electron occupation. We achieve electron densities, lattice distortions, and formation energies in accord with results from piecewise linear hybrid functionals, thereby validating the accuracy of the method. The resulting  energetics is accurate also for polaron hoppings, whereby the use of configurational-dependent $U$ values can be avoided. In this approach, the selection of $U$ is based on physical properties that are directly associated with the orbitals to which $U$ is applied, without involving more global properties, such as band gaps and density of states. For comparison, we also calculate $U$ values through a widely-used linear-response approach \cite{cococcioni2005PRB} finding significantly different values of $U$, which result in a departure from the condition of piecewise linearity. As case studies, we consider  the self-trapped  electron in BiVO$_4$, the self-trapped hole polaron in MgO, the Li-trapped hole in MgO, and the Al-trapped hole trapped  $\alpha$-SiO$_2$.

\section{Results and Discussion}

The DFT+$U$ energy functional can be written in its simplified rotationally-invariant form as \cite{dudarev1998PRB}:
\begin{equation}
	E^{U}[\{\psi_\uparrow^U\},\{\psi_\downarrow^U\}] 
	=   E^0[n_\uparrow^U, n_\downarrow^U]  + \frac{U}{2} \sum_{I\sigma}  \text{Tr} [  \textbf{n}^{I\sigma}(1 - \textbf{n}^{I\sigma})],  \label{eq:dftu_fll_rot_simple}
\end{equation}
where $E^0$ is the semilocal energy, $\psi_{i\sigma}^U$ are the wave functions, $n_\sigma^U = \sum_i |\psi_{i\sigma}^U|$ is the total density, $\sigma$ the spin index, $U$ the Hubbard parameter,  $I$ the atomic site, and  $\mathbf n^{I\sigma}$ the occupation matrix of localized orbitals $\phi_m^I$ of state index   $m$, which is defined as
\begin{equation}
	n_{mm'}^{I\sigma} = \sum_{i} f_{i\sigma} \braket{\psi_{i\sigma}^U | \phi^I_{m}} \! \braket{\phi^I_{m'}| \psi_{i\sigma}^U},
\end{equation}
where $f_{i\sigma}$ are the  occupations of the Kohn-Sham orbitals. Variational minimization of the energy functional $E^{U}$ leads to the following equations:
\begin{equation}
	(\mathcal H_\sigma^0 + V_\sigma^U) \psi_{i\sigma}^U = \epsilon_{i\sigma}^U \psi_{i\sigma}^U,
	\label{eq:ks_dftu}
\end{equation}
where $\mathcal H_\sigma^0$ is the PBE Hamiltonian,  $\epsilon_{i\sigma}^U$ are the eigenvalues, and $V_\sigma^U$ is the DFT+$U$  potential given by \cite{himmetoglu2014JQC}
\begin{equation}
	V_\sigma^U = U \sum_{Imm'}  \bigg[ \frac{\delta_{mm'}}{2} -  n_{mm'}^{I\sigma}   \bigg]  \ket{\phi_{m'}^I}\!\bra{\phi_{m}^I }. \label{eq:pot_U}
\end{equation}
From Eq.\ \eqref{eq:pot_U}, one can see that the Hubbard potential is repulsive for unoccupied orbitals and attractive for occupied orbitals, thereby favoring the Mott localization of electrons on specific atomic sites.

We here consider enforcing the PWL through polaronic defect states associated with the orbitals subject to the correction $U$. The PWL can then be determined nonempirically by finding the value $U=U_\text{k}$ such that the concavity of the total energy upon partial electron occupation vanishes, namely
\begin{equation}
	\left.\frac{d^2}{dq^2} E^U(q)\right|_{U = U_\text{k}}= 0,
	\label{eq:pwl_etot}
\end{equation}
where $q$ is the fractional charge. Through Janak's theorem, the condition in Eq.\ \eqref{eq:pwl_etot} turns into a  constraint on the energy level of the localized state,
\begin{equation}
	\left.\frac{d}{dq} \epsilon_\text{p}^U(q)\right|_{U = U_\text{k}}= 0,
	\label{eq:pwl_ks}
\end{equation}
which requires the energy level to be independent of electron occupation. Equation  \eqref{eq:pwl_ks} can be rewritten as
\begin{equation}
	\frac{d\epsilon_\text{p}^0}{dq} + \frac{d}{dq} \braket{\psi_\text{p}^{U_\text{k}}| V^{U_\text{k}}_\sigma | \psi_\text{p}^{U_\text{k}}} = 0,
	\label{eq:pwl_ks_janak}
\end{equation}
where $\psi_\text{p}^{U_\text{k}}$ is the wave function  of the localized state and $d\epsilon_\text{p}^0/dq$ the variation of the energy level with $q$ as calculated with PBE. We remark that the second term on the right-hand side of  Eq.\ \eqref{eq:pwl_ks_janak} includes complex derivatives of the matrix elements $ n_{mm'}^{I\sigma}$ with respect to $q$. Therefore, it is more practical to determine $U_\text{k}$ by solving Eq.\ \eqref{eq:pwl_ks} by finite differences, namely by imposing that the energy levels calculated at integer charges $q=0$ and $q=Q$ coincide ($Q=-1$ for localized electrons, $Q=+1$ for localized holes). 

For a Hubbard parameter $U$, the  formation energy of the defect state is calculated as \cite{freysoldt2014RMP}
\begin{equation}
	E_\text{f}^U(Q) = E^U(Q) - E_\text{ref}^U(0) + Q \epsilon_\text{b}^U,
	\label{eq:Ef_U}
\end{equation}
where $E^U(Q)$ and  $E_\text{ref}^U(0)$ are the total energies of the defect state and of the reference system, respectively, and $\epsilon_\text{b}^U$ is the relevant band edge of the pristine system. In Eq.\ \eqref{eq:Ef_U}, the defect and reference systems contain the same atoms. We stress that finite-size electrostatic corrections 
due to the use of periodic boundary conditions need to be applied \cite{freysoldt2009PRL,freysoldt2011PSS,komsa2012PRB,falletta2020PRB}. For simplicity of notation, we consider all total energies, formation energies, and energy levels to be corrected by finite-size effects via the expressions in Eqs.\ \eqref{eq:Etot_fs} and \eqref{eq:eps_fs} in Methods.

\begin{figure}[t!]
	\centerline{\includegraphics[width=1.2\linewidth]{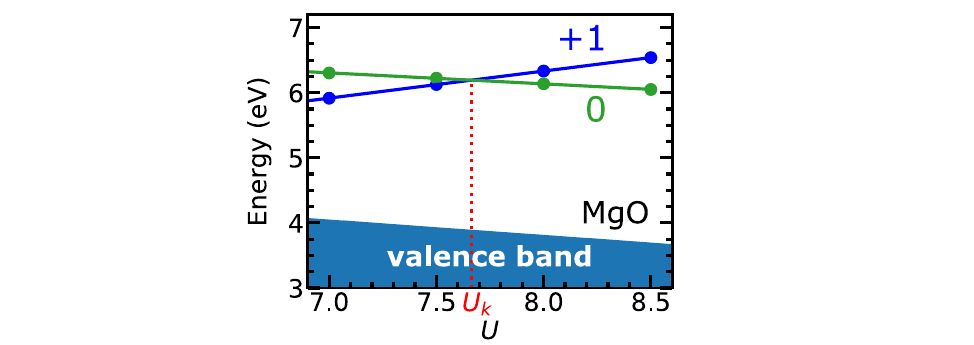}}
	\caption{\textbf{Enforcement of the piecewise linearity.} Energy levels $\epsilon_\text{p}^{U}(+1)$ and $\epsilon_\text{p}^{U}(0)$ as a function of $U$ for the self-trapped hole in MgO. The defect levels are identified by their respective charge. The value $U_\text{k}$ is found such that  $\epsilon_\text{p}^{U_\text{k}}(+1)=\epsilon_\text{p}^{U_\text{k}}(0)$.}
	\label{fig:Uk}
\end{figure}

As case studies, we consider self-trapped and impurity-trapped polaronic defects. In particular, we take the self-trapped electron in BiVO$_4$ \cite{wiktor2018EL}, the self-trapped hole in MgO \cite{varley2012PRB}, the Li-trapped hole in MgO \cite{schirmer1971JPCS,shluger1986JPC}, and the  Al-trapped hole in $\alpha$-SiO$_2$ \cite{pacchioni2000PRB,laegsgaard2001PRL,gerosa2015JCP,varley2012PRB,davezac2005PRB,han2010PRB}.  We remark that, when using the PBE functional, such polaronic states are unstable. Thus, upon structural relaxation, the lattice distortions vanish and the defect charge delocalizes. In particular, self-trapped polarons delocalize over the entire system, and impurity-trapped holes distribute over the O atoms surrounding the impurity. At variance, for the polaronic defects under consideration, DFT+$U$ can stabilize the localized states. We apply the $U$ correction to the orbitals that constitute the localized states, namely the $3d$ orbitals of V atoms in BiVO$_4$, the $2p$ orbitals of O atoms in MgO, and the $2p$ orbitals of O atoms in $\alpha$-SiO$_2$.  In BiVO$_4$, the self-trapped electron localizes on a V atom. In MgO, the self-trapped hole localizes on a O atom. In Li-doped MgO, the hole localizes on a O atom neighbouring the Li site.  In $\alpha$-SiO$_2$, the hole localizes on a O atom neighbouring the Al site.  Additional computational details are given in Methods.

We  determine the Hubbard parameter $U_\text{k}$ through the enforcement of Eq.\ \eqref{eq:pwl_ks}.  We proceed as follows. We obtain the defect structure  at various values of $U$ by performing self-consistent structural relaxations. At such fixed structures, we calculate the energy levels $\epsilon_\text{p}^{U}(Q)$ and $\epsilon_\text{p}^{U}(0)$ accounting for finite-size effects [cf.\ Eq.\ \eqref{eq:eps_fs} in Methods]. By imposing 
that $\epsilon_\text{p}^{U_\text{k}}(Q) = \epsilon_\text{p}^{U_\text{k}}(0)$, we then obtain $U_\text{k} = 3.5$, 7.7, 7.5, and 8.3 eV  for the self-trapped electron in BiVO$_4$, the self-trapped hole in MgO, the Li-trapped hole in MgO, and Al-trapped hole in $\alpha$-SiO$_2$, respectively. This procedure is illustrated in Fig.\ \ref{fig:Uk} for the self-trapped hole in MgO. We remark that the values of $U_\text{k}$ obtained for the self-trapped and the Li-trapped holes in MgO differ by only 0.2 eV, indicating that our scheme is robust upon varying the polaronic defect. 
This is analogous to the case of hybrid functionals, where one observes a weak dependence of $\alpha_\textrm{k}$ on the defect used for enforcing the PWL \cite{bischoff2019PRB,bischoff2019PRM,bischoff2021PRR,miceli2018PRB}. In this context, we remark that  finite-size corrections crucially affect the value of $U_\text{k}$. Indeed, without such corrections, we would have obtained $U_\text{k}^\text{uncor}=$ 1.7, 4.9, 4.6, 5.1 eV for our respective case studies, with \text differences with respect to the corrected values amounting up to 3.2 eV. This emphasizes the importance of correcting for finite-size errors.

\begin{figure}[t!]
	\centerline{\includegraphics[width=1.2\linewidth]{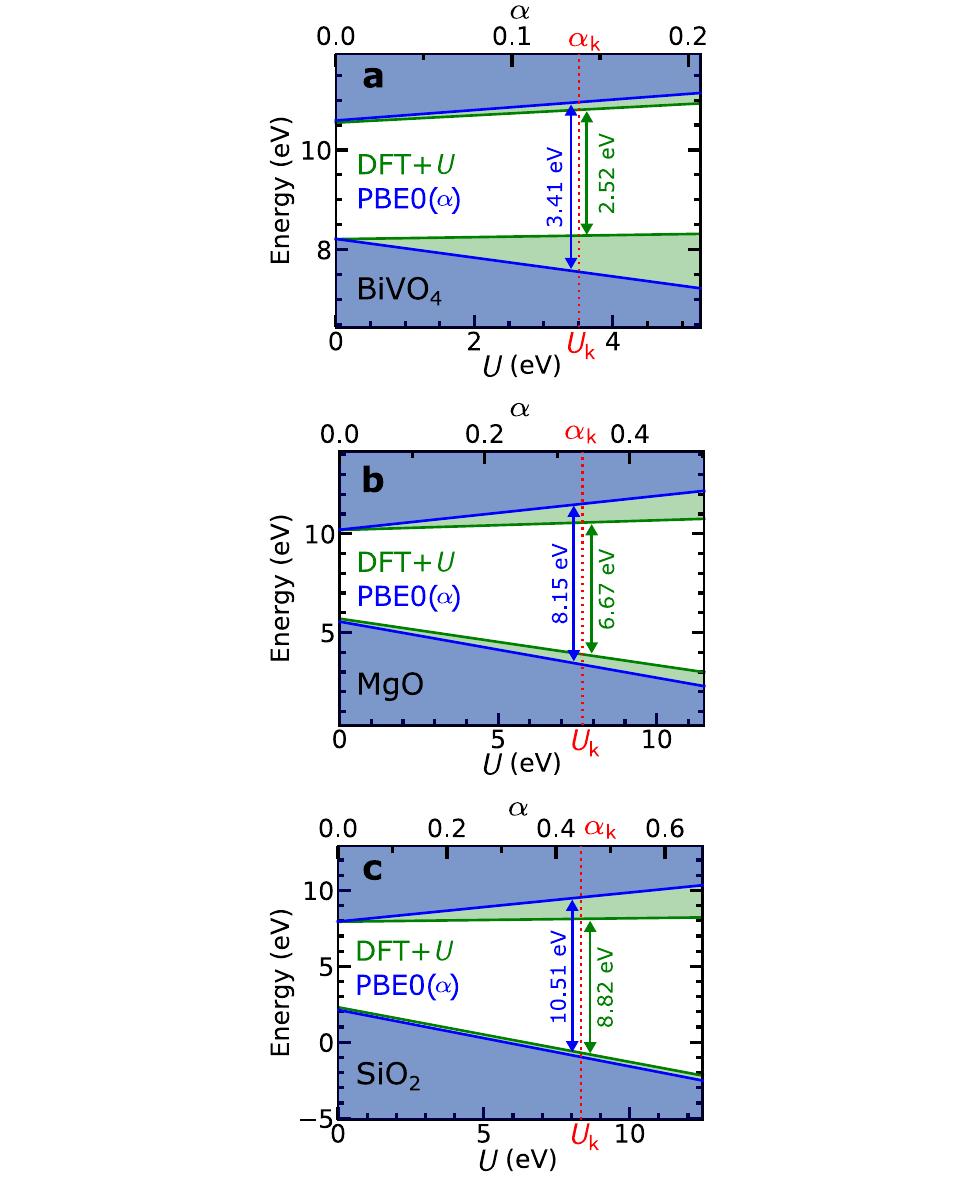}}
	\caption{\textbf{Band gaps obtained with various functionals.} Band edges as obtained with DFT+$U$ and PBE0($\alpha$) as a function of $U$ and $\alpha$, respectively, for BiVO$_4$, MgO, and $\alpha$-SiO$_2$.  For MgO, we consider $U_\text{k}$ and $\alpha_\text{k}$ calculated for the self-trapped hole. The vertical red line denotes the choice of the parameter for which the piecewise linearity condition is retrieved. The energy levels are aligned with respect to the average electrostatic potential \cite{Alkauskas2011PRB}.}
	\label{fig:bandgap}
\end{figure}

\begin{table}[t!]
	\caption{\textbf{Band gaps.} Band gaps calculated with DFT+$U_\text{k}$ ($E_\text{g}^{U_\text{k}}$) and PBE0($\alpha_\text{k}$) ($E_\text{g}^{\alpha_\text{k}}$) compared to reference experimental values  after adding appropriate corrections ($\Delta E_\text{g}$)  taken from Ref.\ \cite{falletta2022PRB}. The reference experimental values correspond to the optical band gap at 300 K for BiVO$_4$ \cite{sayama2006JPCB,lyo2008JPCC,kudo1999JACS}, the fundamental band gap at 6 K for MgO \cite{onuma2021APL}, and the first peak in the reflectance spectrum for $\alpha$-SiO$_2$. In MgO, $U_\text{k}$ is obtained from the self-trapped hole. Energies are in eV. }
	\label{tab:band_gaps}
	\begin{ruledtabular}
		\begin{tabular}{l *{5}{d{1.3}} l }
			& \mc{$E_\text{g}^{U_\text{k}}$} & \mc{$E_\text{g}^{\alpha_\text{k}}$}  &  \mc{$\Delta E_\text{g}$} & \mc{$E_\text{g,cor}^{U_\text{k}}$} & \mc{$E_\text{g,cor}^{\alpha_\text{k}}$}  & \mc{Expt.} \\  \hline \rule{-4pt}{3ex}
			BiVO$_4$          & 2.52 & 3.41  & -1.16 & 1.36 & 2.25 & 2.4-2.5  \\ \rule{-4pt}{3ex} 
			MgO                       & 6.67 & 8.15 & -0.53 &  6.14 & 7.62& 7.77  \\  \rule{-4pt}{3ex}
			$\alpha$-SiO$_2$  & 8.82 & 10.51  & 0.02 & 8.84 & 10.53 & 10.30   \rule{-4pt}{3ex}
		\end{tabular}
	\end{ruledtabular}
\end{table}

It is of interest to investigate the band gaps resulting from our selection of $U$. In Fig.\ \ref{fig:bandgap}, we show the evolution of the band gaps obtained with DFT+$U$ as function of $U$ and of the band gaps obtained with PBE0($\alpha$) as a function of $\alpha$. In correspondence of $U_\textrm{k}$, DFT+$U$ yields
band gaps of 2.52, 6.67, and 8.82 eV for BiVO$_4$, MgO, and $\alpha$-SiO$_2$, respectively. For MgO, we here use the value of $U_\text{k}$ calculated for the self-trapped hole, considering the negligible difference with respect to the value for the Li-trapped hole. 
After the inclusion of appropriate corrections due to spin-orbit coupling, phonon renormalization, and exciton binding energies \cite{falletta2022PRB}, the DFT+$U_\textrm{k}$ band gaps are found to noticeably differ from their experimental counterparts (cf.\ Table \ref{tab:band_gaps}).
These discrepancies contrast with the case of hybrid functionals, for which the agreement with experiment is 
within 0.25 eV (cf.\ Table \ref{tab:band_gaps}). The good performance of hybrid functionals derives from a global improvement of the electronic structure, in accord with numerous previous studies 
\cite{deak2017PRB,miceli2018PRB,bischoff2019PRB,bischoff2019PRM,wing2020PRM,wing2021PNAS,yang2022JPCL,smart2018PRM}.
From this analysis, we infer that an accurate description of band gaps should generally not be expected from DFT+$U_\text{k}$. We assign this to the fact that the $U$ correction only applies to a subset of orbitals, which are not necessarily involved in both valence and conduction bands.
Nevertheless, we expect that physical properties directly associated with the $U$-corrected orbitals should be properly described in DFT+$U_\text{k}$. For instance, in the case of polaronic defects, the formation energies express the relative stability of localized and delocalized states both being constituted by the same $U$-corrected orbitals.
 
 \begin{figure}[b!]
 \centerline{\includegraphics[width=1.0\linewidth]{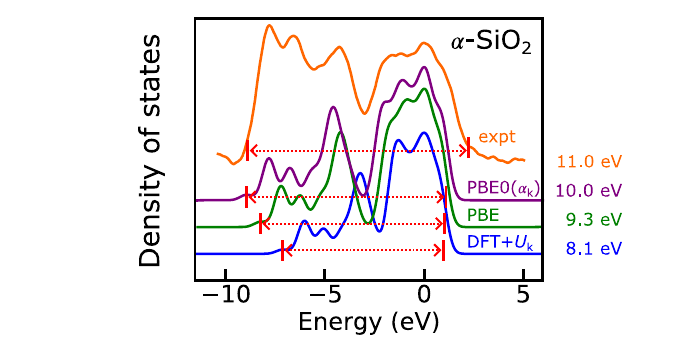}}
 \caption{\textbf{Band widths obtained with various functionals.} Density of valence band states for $\alpha$-SiO$_2$ as calculated with PBE0($\alpha_\text{k}$), PBE, and DFT+$U_\text{k}$, compared with the experimental XPS spectrum from Ref.\ \cite{laughlin1979PRB}. The corresponding band widths are indicated. The theoretical band widths correspond to differences between Kohn-Sham levels, whereas the experimental band width is obtained from  extrapolations of the wings. The curves are aligned with  respect to the position of the highest energy peak. }
 \label{fig:dos}
\end{figure}

Similar arguments apply when considering the effect of the Hubbard parameter $U_\text{k}$ on the density of states. As test case, we take  $\alpha$-SiO$_2$ and compare the density of states obtained with DFT+$U_\text{k}$  with respect to experiment. As illustrated in Fig.\ \ref{fig:dos},  DFT+$U_\text{k}$ yields a valence band width of 8.1 eV, which is lower than  both the corresponding PBE value of 9.3 eV and the experimental value of 11.0 eV \cite{laughlin1979PRB}. This confirms the common finding that DFT+$U$ narrows the band widths \cite{imada1998RMP}. Hence, in analogy to our discussion on band gaps, DFT+$U_\text{k}$ should not be expected to reproduce more global properties such as the density of states,  even though the polaronic properties are reasonably well captured.  This should be contrasted with the case of the hybrid functional  PBE0($\alpha_\text{k}$), where the globally improved functional also yields an improved band width. Indeed, in the case of $\alpha$-SiO$_2$, we find a PBE0($\alpha_\text{k}$) band width of 10.0 eV, improving upon the PBE value of 9.3 eV (cf.\ Fig.\ \ref{fig:dos}).

\begin{figure*}[t!]
	\centerline{\includegraphics[width=1.0\linewidth]{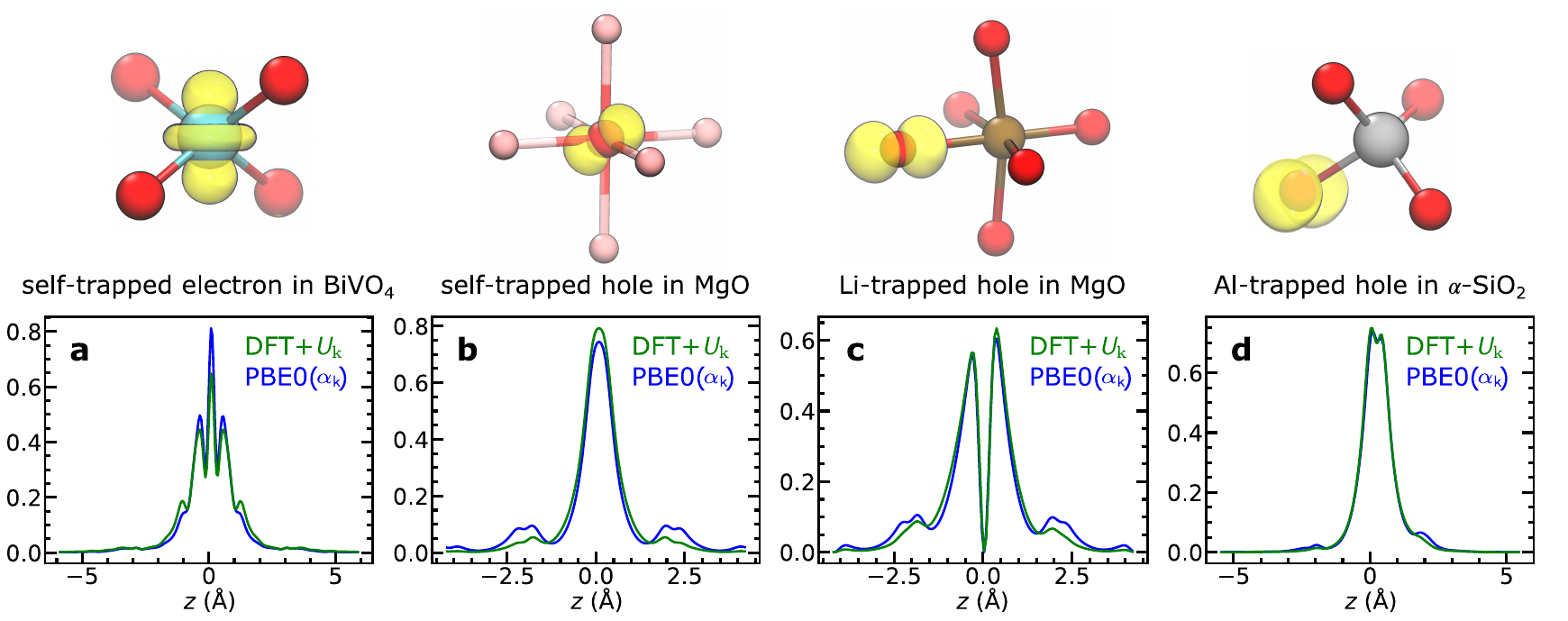}}
	\caption{\textbf{Polaron electron density.} Electron densities obtained with DFT+$U_\text{k}$ and PBE0($\alpha_\text{k}$) functionals for the self-trapped electron in BiVO$_4$, the self-trapped hole in MgO, Li-trapped hole in MgO, and the Al-trapped hole in $\alpha$-SiO$_2$.  The defect density is integrated over $xy$-planes. On top, isodensity surfaces at 5\% of their maximum (Bi in orange, V in cyan, O in red, Mg in pink, Li in brown, Si in blue, Al in grey).}
	\label{fig:density}
\end{figure*}

We calculate electronic and structural properties of the polaronic defects studied in this work using the DFT+$U_\text{k}$ functional and compare the results with those from PBE0($\alpha_\text{k}$) hybrid functionals. 
Details of the hybrid functional calculations are given in Methods.
As illustrated in  Fig.\ \ref{fig:density}, we find very good agreement between the defect densities calculated with the two schemes. Moreover, the lattice distortions practically coincide, with bond lengths deviating by at most $0.03~\text{\AA}$ (cf.\ Table \ref{tab:structures}). Using Eq.\ \eqref{eq:Ef_U}, we calculate the respective formation energies  $E_\text{f}^{U_\text{k}} = -0.49$, $-$0.64, $-2.01$, and $-$3.27 eV. These values are given in Table \ref{tab:Ef_U}.  Deviations from PBE0($\alpha_\text{k}$) results amount to at most 0.19 eV (cf.\ Table \ref{tab:Ef_U}). This extends the robustness of piecewise linear functionals to DFT+$U$ schemes  \cite{falletta2022PRL,falletta2022PRB}, and concurrently validates our criterion for determining the value of $U$.

\begin{table}[t!]
	\caption{\textbf{Polaron structure.} Bond lengths (in \AA) of the polaronic defect structures  obtained with DFT+$U_\text{k}$ and PBE0($\alpha_\text{k}$) functionals.  For the Li-trapped hole in MgO, we give the lengths of the short/intermediate/long Li-O bonds. For the Al-trapped hole in $\alpha$-SiO$_2$, we give the lengths of the short/long Al-O bonds.}
	\label{tab:structures}
	\begin{ruledtabular}
		\begin{tabular}{l *{2}{d{1.3}} }	
			Polaronic defect & \mc{DFT+$U_\text{k}$ } & \mc{PBE0($\alpha_\text{k}$)} \\  \hline \rule{-4pt}{3ex}
			BiVO$_4$   (self-trapped)           &  1.82   &  1.80 \\  \rule{-4pt}{3ex} 
			MgO            (self-trapped)          &  2.22   & 2.20 \\  \rule{-4pt}{3ex}
			MgO            (Li-trapped)   &  \mc{1.92/2.17/2.30}   & \mc{1.90/2.17/2.33} \\  \rule{-4pt}{3ex}
			$\alpha$-SiO$_2$ (Al-trapped)  &  \mc{1.67/1.92} & \mc{1.69/1.91}  \rule{-4pt}{3ex}
		\end{tabular}
	\end{ruledtabular}
\end{table}	

 \begin{figure}[b!]
	\centerline{\includegraphics[width=1.1\linewidth]{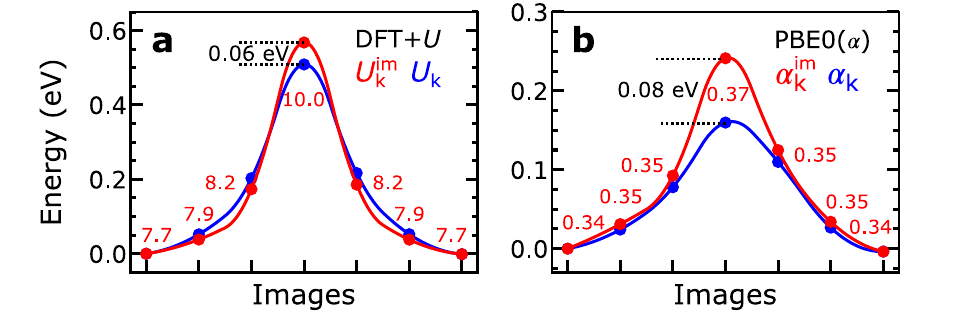}}
	\caption{\textbf{Polaron hopping barriers obtained with various functionals.} Energy along a polaron pathway connecting two neighbouring O atoms in MgO, as calculated (\textbf{a}) with  fixed $U_\text{k}$ and image-dependent $U_\text{k}^\text{im}$ in DFT+$U$ calculations, and (\textbf{b}) with fixed $\alpha_\text{k}$ and image-dependent $\alpha_\text{k}^\text{im}$ in PBE0($\alpha$) calculations. The values of $U_\text{k}^\text{im}$ and $\alpha_\text{k}^\text{im}$ for individual images are given.}
	\label{fig:polaron_neb_MgO}
\end{figure}

We further investigate the accuracy of the DFT+$U_\text{k}$ energetics along a polaron hopping pathway. As test case, we consider the hopping  of a hole polaron between two neighboring sites in MgO. We construct a 7-image migration pathway through linear interpolation of the initial and final states. First, we evaluate the energy along the path using the determined value of $U_\text{k}$, as given in Fig.\ \ref{fig:polaron_neb_MgO}(a). Next,  we determine $U_\text{k}^\text{im}$ through the enforcement of the PWL for each image, finding the largest deviation with respect to  $U_\text{k}$  in correspondence of the transition state. This is due to the fact that at the transition state the polaron density is equally distributed among two neighboring O sites, thus deviating the most from the case of the hole polaron trapped at a single O site.  Then, we calculate the energy along the pathway as $E^{U_\text{k}^\text{im}}[\text{polaron,im}] - E^{U_\text{k}^\text{im}}[\text{bulk}]$ for each image. As illustrated in Fig.\ \ref{fig:polaron_neb_MgO}(a), the difference between the energy barriers calculated with either fixed $U_\text{k}$ or image-dependent  $U_\text{k}^\text{im}$ amounts to only 0.06 eV. This validates the choice of a fixed $U_\text{k}$ for polaron hopping calculations. We carry out the same analysis with the PBE0($\alpha$) hybrid functional,  finding a difference of 0.08 eV between the barriers calculated with either fixed $\alpha_\text{k}$ or image-dependent $\alpha_\text{k}^\text{im}$ [cf.\ Fig.\ \ref{fig:polaron_neb_MgO}(b)]. The energy barriers obtained with DFT+$U_\text{k}$ and PBE0$(\alpha_\text{k})$ differ by 0.32 eV, which is comparable with the typical accuracy achieved upon enforcing the PWL with different functionals (cf.\ Table  \ref{tab:Ef_U} and Refs.\ \cite{falletta2022PRL,falletta2022PRB}).

For comparison, we also determine $U$ using the linear-response approach introduced by Cococcioni and de Gironcoli \cite{cococcioni2005PRB}. In this method, the parameter $U$ is chosen to enforce the PWL in density-functional perturbation theory as 
\begin{equation}
	U_\text{\sc lr} = (\chi_0^{-1}  - \chi^{-1})_{II},
	\label{eq:U_lr}
\end{equation}
where  $\chi$  and $\chi_0$  are  screened and unscreened response matrices, respectively, which are defined as variations of the occupations $n^I=\sum_{\sigma m} n_{mm}^{I\sigma} $ with respect to perturbations $\alpha^J$ of the electronic occupations at site $J$. We determine $U_\text{\sc lr}$ on neutral bulk structures using the PBE wave functions. 
We find $U_\text{\sc lr} = 5.4, 10.9, 10.1$ eV for BiVO$_4$, MgO, and $\alpha$-SiO$_2$, respectively. The resulting formation energies of the polaronic defects studied in this work are $E_\text{f}^{U_\text{\sc lr}} = -$1.34, $-$1.67, 3.09, and $-$4.00 eV, as given in Table \ref{tab:Ef_U}. We remark that $U_\text{\sc lr}$ is noticeably larger than $U_\text{k}$  in all cases. Consequently, the  respective formation energies calculated with $U_\text{k}$ and $U_\text{\sc lr}$ differ by 0.85, 1.03, 1.08, and 0.73 eV. These large variations are in part due to the shift of the band edges upon variation of $U$ (cf.\ Fig.\ \ref{fig:bandgap}), which enter in the definition of the formation energy in Eq.\ \eqref{eq:Ef_U}. To assess the dependence on the adopted configuration in the context of this comparison, we also use the linear-response approach on the very same polaron configuration used for the determination of $U_\text{k}$ in the direct piecewise linear scheme. In this way, the same configurational set-up is used in the two approaches, thereby enabling a direct comparison. We take the $U_\textsc{lr}'$ parameter resulting from the linear-response scheme for the atom where the polaron is localized. Focusing on the hole polaron in MgO,  we find $U_\textsc{lr}' = 9.4$  eV, to be compared with the respective value $U_\text{k}=7.7$ eV found through the direct application of the PWL condition. Thus, this analysis further confirms that the structural configuration is not at the origin of the differences between the two schemes for the determination of $U$. Additionally, we remark that our $U_\text{k}$ is found for a Hubbard correction acting on all the atoms of the same species at the same time, whereas in the linear-response approach $U_\textsc{lr}$ is found through a variation on a single atom. Hence, for an even closer comparison, we also determine the value $U_\text{k}'$ by enforcing the PWL upon the application of $U$ to the sole atom where the polaron localizes. In the case of the hole polaron in MgO, we find $U_\text{k}' = 8.5$ eV, which still differs sizably from $U_\textsc{lr}'=9.4$ eV. This further confirms that the differences between the two methods are not only related to the computational setup.

\begin{table}[t]
	\caption{\textbf{Comparison between different schemes for the determination of \textit{\textbf{U}}.} Hubbard parameter $U_\text{k}$ obtained with the scheme introduced in this work compared with the parameter $U_\text{\sc lr}$ resulting from the linear-response method  \cite{cococcioni2005PRB}, together with the corresponding defect formation energies. For reference, we also give the formation energies $E_\text{f}^{\alpha_\text{k}}$ obtained with the piecewise linear PBE0($\alpha_\text{k}$) hybrid functional.}
	\label{tab:Ef_U}
	\begin{ruledtabular}
		\begin{tabular}{l *{5}{d{1.3}} }	
			Defect  & \mc{$U_\text{k}$} & \mc{$U_\text{\sc lr}$} & \mc{$E_\text{f}^{U_\text{k}}$} & \mc{$E_\text{f}^{U_\text{\sc lr}}$} & \mc{$E_\text{f}^{\alpha_\text{k}}$} \\  \hline \rule{-4pt}{3ex}
			BiVO$_4$ (self-trapped)              &  3.5  &  5.4 & -0.49 &  -1.34 & -0.63  \\  \rule{-4pt}{3ex} 
			MgO  (self-trapped)                     &  7.7  &  10.9 & -0.64  & -1.67 & -0.53  \\  \rule{-4pt}{3ex}
			MgO (Li-trapped)               &  7.5  & 10.9  &  -2.01  & -3.09  & - 1.82	  \\  \rule{-4pt}{3ex}
			$\alpha$-SiO$_2$ (Al-trapped)  &  8.3  &  10.1  & -3.27 &  -4.00 & -3.11 \rule{-4pt}{3ex}
		\end{tabular}
	\end{ruledtabular}
\end{table}	

\begin{figure}[t!]
	 \centerline{\includegraphics[width=1.05\linewidth]{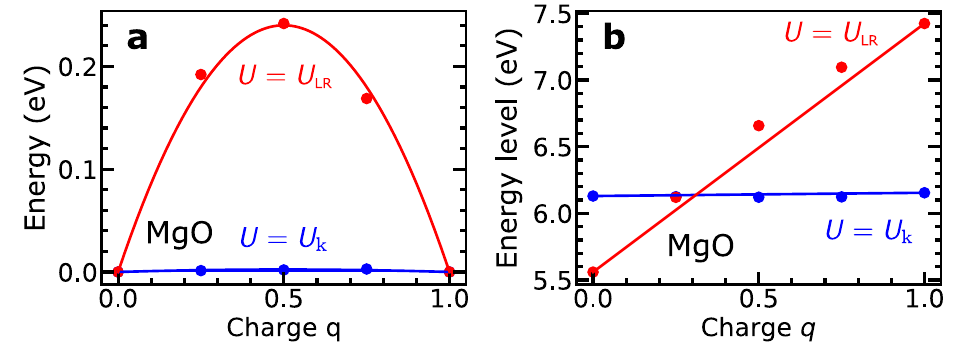}}
	\caption{\textbf{Piecewise linearity of different schemes for the determination of \textit{\textbf{U}}}. (\textbf{a}) Deviation from the piecewise linearity of the total energy and (\textbf{b}) dependence of the defect level on the charge $q$, for the self-trapped hole in MgO. Results for $U_\textrm{k}$ and $U_\text{\sc lr}$ are compared. The solid lines are a guide to the eye.}
	\label{fig:pwl}
\end{figure}

The significant differences between $U_\text{k}$ and $U_\text{\sc lr}$ call for a deeper investigation. Since both approaches are designed to enforce the PWL, we explicitly verify the extent by which the PWL is satisfied in the two schemes. This can be achieved by studying the total energy and the defect level as a function of $q$ for the two choices of the parameter $U$. As illustrated in Fig.\ \ref{fig:pwl},  $U_\text{k}$ indeed yields a piecewise linear total energy and a constant defect level with respect to partial electron occupation. At variance, for $U_\text{\sc lr}$,  the total energy is convex with $q$, and the defect level is not constant. To  understand these differences, we remark that the Kohn-Sham equations used to determine $U$ in the linear-response approach are
\begin{equation}
	\Big(\mathcal H_\sigma^0 + \alpha^I \sum_m \ket{\phi_m^I}\!\bra{\phi_m^I}\Big) \psi_{i\sigma}^{\alpha^I} = \epsilon_{i\sigma}^{\alpha^I} \psi_{i\sigma}^{\alpha^I}, \label{eq:ks_DFPT}
\end{equation}
where $\alpha^I$ is the amplitude of the perturbation, and $\epsilon_{i\sigma}^{\alpha^I}$ and $\psi_{i\sigma}^{\alpha^I}$ are the corresponding eigenvalues and wave functions.  The Hamiltonian in Eq.\ \eqref{eq:ks_DFPT} differs from the  DFT+$U$ Hamiltonian in  Eq.\ \eqref{eq:ks_dftu}, whereby the $U$ values that enforce the PWL in the two cases could be different. This could underlie the departure from the PWL observed in Fig.\ \ref{fig:pwl} for DFT+$U_\text{\sc lr}$. However, we remark that despite the different defect formation energies, the electron densities and the structural distortions of the polaronic defects obtained with $U_\text{k}$ and $U_\text{\sc lr}$ practically coincide. 

In conclusion, our work addresses the determination of the Hubbard $U$ in the DFT+$U$ functional through enforcing the piecewise linearity condition on polaronic defect states. Our selection of $U$ yields electronic and structural properties of such defects in good agreement with results from hybrid functionals satisfying the same constraint.  
Our scheme is further validated by the excellent agreement found for formation energies obtained with piecewise linear functionals.  We demonstrate that our criterion for $U$ leads to accurate energy barriers in polaron hoppings, whereby configurational-dependent $U$ values can be avoided. We emphasize that our approach targets physical properties related to the $U$-corrected orbitals, while more global properties, such as band gaps and band widths, are not directly involved.  For comparison, we also calculate $U$ through a widely-used linear-response method, finding values of $U$ that break the piecewise linearity condition and give larger formation energies.  To sum up, we showed that polaronic defect states can  effectively be used for determining the value of the Hubbard $U$ parameter in DFT+$U$. Additionally, we demonstrated that the resulting electronic, structural, and 
energetic properties of such defects closely correspond to those obtained with hybrid functionals, but at a 
noticeably lower computational cost.

\subsection{Methods}
\footnotesize 

\noindent \textbf{Computational details}.  The calculations are performed using the version 7.1 of the \textsc{quantum espresso} suite \cite{giannozzi2009JPCM}.  The core-valence interactions are described by normconserving pseudopotentials \cite{vansetten2018CPC}.  BiVO$_4$ is modeled with a 96-atom orthorhombic supercell ($a = 10.34$ \AA, $b=10.34$ \AA, $c = 11.79$ \AA), MgO  with a 64-atom cubic supercell ($a = 8.45$ \AA), and $\alpha$-SiO$_2$ with a 72-atom hexagonal supercell ($a = 9.97$ \AA, $c = 10.96$ \AA).  We optimize the lattice parameters and the atomic positions using the PBE functional for the pristine systems. The Brillouin zone is sampled at the $\Gamma$ point and the energy cutoff is set to 100 Ry in all cases.  We obtain the electron and hole polarons by either adding or removing one electron, respectively. The defect structures are  relaxed at fixed supercell parameters. The high-frequency and static dielectric constants  used for the determination of the finite-size effects \cite{falletta2020PRB} are  calculated by applying finite electric fields \cite{umari2002PRL} at the semilocal level of theory \cite{falletta2022PRL,falletta2022PRB}. The Hubbard parameters $U_\text{\sc lr}$ are calculated using the code \textsc{hp}  \cite{timrov2022CPC}.  \\

\noindent \textbf{Hybrid functional calculations}. 
The procedure for determining  $\alpha_\text{k}$ is analogous to that for $U_\text{k}$ (see Refs.\ \cite{falletta2022PRL,falletta2022PRB}). 
The hybrid functional results for the self-trapped polarons in BiVO$_4$ and MgO, and the Al-trapped hole in $\alpha$-SiO$_2$ are taken from Refs.\ \cite{falletta2022PRL,falletta2022PRB}, in which the same computational setup has been employed. For the Li-trapped hole in MgO, we obtain $\alpha_\text{k} = 0.33$, which is in good agreement with the value $\alpha_\text{k} = 0.34$ found for the self-trapped hole \cite{falletta2022PRL,falletta2022PRB}. The corresponding formation energy is $-1.82$ eV and is given in Table \ref{tab:Ef_U}.
 \\

\noindent \textbf{Finite-size corrections}.  For a system with supercell charge $q^*$ in a geometry $\mathbf R_{Q^*}$, relaxed in the presence of a charge $Q^*$, the finite-size correction for the total energy is given by \cite{falletta2020PRB}
\begin{align}
	E_\text{cor}(q^*,\mathbf R_{Q^*}) & = E_\text{m}(Q^*,\varepsilon_0) - E_\text{m}(Q^* + Q^*_\text{pol},\varepsilon_\infty) \nonumber \\
	& \quad + E_\text{m}(q^*+Q^*_\text{pol},\varepsilon_\infty), \label{eq:Etot_fs}
\end{align}
where $E_\text{m}$ denotes the finite-size correction  for defects screened through either the high-frequency ($\varepsilon_\infty$) or the static ($\varepsilon_0$) dielectric constant \cite{freysoldt2009PRL,freysoldt2011PSS},  and $Q^*_\text{pol}=-Q^*(1-\varepsilon_\infty/\varepsilon_0)$ is the ionic polarization charge associated with the frozen lattice distortions. Through Janak's theorem, the corresponding finite-size correction for the defect energy level is \cite{falletta2020PRB}
\begin{equation}
	\epsilon_\text{cor}(q^*,\mathbf R_{Q^*}) = - 2 \frac{E_\text{m}(q^*+Q^*_\text{pol},\varepsilon_\infty)}{q^*+Q^*_\text{pol}}. \label{eq:eps_fs}
\end{equation}
 We remark that the supercell charges $q^*$ and $Q^*$ coincide with the polaron charges $q$ and $Q$ for self-trapped polarons, as in BiVO$_4$ and MgO. At variance, in the cases of Li-doped MgO and Al-doped $\alpha$-SiO$_2$, $q^*=q-1$ and $Q^*=Q-1$ since the hole trapping occurs in the neutral state.

\subsection{Data availability}

The data associated with this work can be found on Materials Cloud \cite{cloud}.

\subsection{Ackowledgements}
The calculations have been performed at the Swiss National Supercomputing Centre (CSCS) (grant under Projects ID s1122). 

\subsection{Author contributions}

Both authors conceived the project. S.\ F.\ performed the numerical calculations. Both authors contributed to the writing of the manuscript.

\subsection{Competing interests}

The authors declare no competing interests.

\bibliography{bibliography}

\end{document}